\documentstyle[aps,prl,graphicx]{revtex}

\begin{document} 
\twocolumn[\hsize\textwidth\columnwidth\hsize\csname @twocolumnfalse\endcsname \vskip1pc] \narrowtext 
\noindent{\bf \mbox{Comment on ``Strong Vortex Liquid Correlation} from Multiterminal Measurements on Untwinned  YBa$_2$Cu$_3$O$_{7-\delta}$ Single Crystals''}\\
From recent transport experiments 
%using a pseudo-transformer configuration  
A.Rydh and \"O.Rapp [1] (RR) claim that the vortex liquid in untwinned YBa$_2$Cu$_3$O$_{7-\delta}$ crystals is correlated above the melting transition, in striking contrast to previous work [2]. 
In this Comment we present new measurements using the same experimental technique on twinned and untwinned YBa$_2$Cu$_3$O$_{7-\delta}$ crystals with similar overall characteristics as those reported by RR. The comparison of the vortex correlation response in both cases indicates that the central conclusion of the work of RR is not correct. Our results reconfirm the work by L\'opez {\it et al.} [2] and points on the origin of the misinterpretation in the work of RR.

The results of the voltage response, V(T), for a {\it twinned} crystal in a magnetic field of H = 6 T, are shown Fig.1(a). At high temperatures the dissipation at the top, V$_{top}$, is larger than at the bottom, V$_{bot}$, as a result of the inhomogeneous current distribution acting upon an uncorrelated vortex system. As the temperature is lowered V$_{top}$ equals V$_{bot}$  at T$_{th}$. This result could be either  a trivial consequence of the limited experimental voltage sensitivity (indicating an ohmic decrease of the resistance in the $c$ direction) or associated with the stablishment of vortex velocity correlation wich implies vortex phase
correlation along the field direction across the sample thickness. In the first case the I-V curves are ohmic, in the second a non-linear behavior indicates the existence of a finite cutting current [3], I$_{cut}$, for T$<$T$_{th}$. This second case is fulfilled for the twinned sample as shown in the inset of Fig. 1(a). While for T$>$T$_{th}$,  $\Delta$V(I)= V$_{top}$-V$_{bot}$ extrapolates to zero at I=0, for T$<$T$_{th}$ a well defined I$_{cut}$ is detected.

For an {\it untwinned} crystal [4] equivalent results are shown in Fig. 1(b). From the V(T) curves alone one is tempted to conclude, as RR did, that the vortex liquid {\it is correlated} between T$_m$ and the ``apparent'' T$_{th}$. However, the behavior of the I-V curves shows that in this case the result V$_{top}$=V$_{bot}$ is just a  consequence of the limited voltage sensitivity. The inset shows a linear $\Delta$V(I) that extrapolates to zero for I=0 at $T < ``T_{th}"$, consistent with a continuous ohmic decrease of the resistance in the $c$ direction. 
This absence of I$_{cut}$ is in sharp contrast to the behavior observed in the twinned samples [2] and is the fingerprint of an {\it uncorrelated} vortex system. Moreover, the local analysis of the parallel and perpendicular resistivities in terms of the Laplace equation [1] is only consistent with this linear behavior [5].

In summary, the information provided by the I-V curves is crucial to give a correct interpretation to the T$_{th}$ extracted from the V(T) curves. This key point is lacking in the results of RR and therefore it cannot be concluded solely from their V(T) curves that the vortex liquid in untwinned samples is correlated. A proper measurement of I-V characteristics would have led RR to the opposite conclusion.
%the vortex liquid above the melting temperature in untwinned crystals is indeed uncorrelated in the field direction, as we have shown. 

We thank G. Nieva for the x-ray and polarized light studies and Y. Fasano for the Bitter decoration.\\
 \noindent
 \vspace{.2cm}\\
B.Maiorov, A.V.Silhanek, F. de la Cruz and E.Osquiguil\\
\\
\hspace{1cm}{\small Centro At\'{o}mico Bariloche and Instituto Balseiro, 
Comisi\'on Nacional de Energ\'{\i}a At\'omica, 8400 S. C. de Bariloche, R. N., Argentina.
 \vspace{.2cm}\\
$[1]$ A.Rydh and \"O.Rapp, Phys. Rev. Lett. {\bf 86}, 1873 (2001).\\
$[2]$ D.L\'opez {\it et al.}, Phys. Rev. Lett. {\bf 76}, 4034 (1996).\\
$[3]$ D.L\'opez {\it et al.}, Phys. Rev. B {\bf 50}, 9684 (1994).\\
$[4]$ as confirmed by x-rays, polarized light microscopy and Bitter decoration of the vortex lattice.\\
$[5]$ H.Safar {\it et al.}, Phys. Rev. Lett. {\bf 72}, 1272 (1994).\\
}

\begin{figure}[htb]
\begin{center}
\includegraphics[width=75mm]{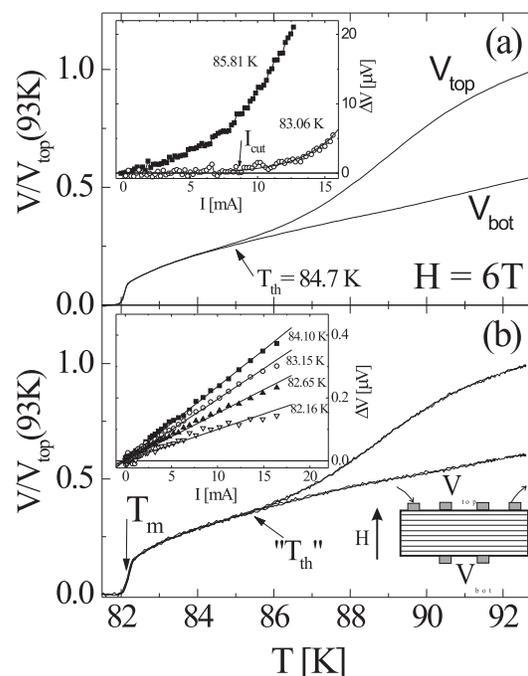}
\caption{Normalized V(T) measurements for:(a) Twinned crystal. Inset: $\Delta V(I)$ curves for $T>T_{th}$ and $T<T_{th}$. (b) Untwinned crystal. Inset: $\Delta V(I)$ curves for $T<``T_{th}"$.}
\end{center}
\label{figure}
\end{figure}

\end{document}